\documentclass[english]{aastex}
\usepackage[T1]{fontenc}
\usepackage{float}
\usepackage{amstext}
\usepackage{graphicx}

\makeatletter

\newcommand*\LyXbar{\rule[0.585ex]{1.2em}{0.25pt}}

\slugcomment{}
\shorttitle{}
\shortauthors{}

\makeatother

\usepackage{babel}

\begin{document}

\title{The Adams-Bashforth-Moulton Integration Methods Generalized to an
Adaptive Grid}

\author{A. P. Hayes}

\affil{University of Maine, 120 Bennett Hall, Orono, Maine 04469-5709, USA}

\email{andrew.hayes@umit.maine.edu}
\begin{abstract}
We present a generalization of the Adams-Bashforth-Moulton predictor-corrector
numerical integration methods to an adaptive grid. The step size may
be chosen dynamically in order to maintain a desired relative magnitude
of error in each step. We demonstrate that the methods remain convergent
to the expected degree, and apply various methods to the famous problem
of determining the maximum possible mass of a neutron star supported
by pure fermionic exclusion pressure. We reproduce the Tolman-Oppenheimer-Volkoff
result of 0.71 solar masses using only 23 integration steps, and reproducing
both mass and radius within 1\% requires 27. We also present various
optimizations and features of our implementation.
\end{abstract}


\section{INTRODUCTION\label{sec:Introduction}}

The Adams-Bashforth (AB) family of integration methods \citep{BA}
are explicit, linear, multistep techniques. Each successive member
of the family has a higher order of convergence, and the family can
be extended indefinitely. The Adams-Moulton (AM) family of integration
methods \citep{Moulton} are, similarly, implicit, linear, multistep
techniques, and can be similarly extended to arbitrarily high order
of convergence. Combining the two allows the use of an Nth order,
AB method's explicit result for an integration step as a prediction
to be inserted into the AM method of order N+1, thereby achieving
a next-order correction for the integration step, as well as an estimate
of the error in that step. We will term this predictor-corrector combined
method Adams-Bashforth-Moulton. For clarity, we will refer to the
orders of convergence of both the Adams-Bashforth predictor phase
and the Adams-Moulton correction phase, e.g. {}``ABM fixed-grid method
of order 3-4.''

The principle of traditional, fixed-grid AB and AM methods is to integrate
analytically a Lagrange interpolating polynomial fit to various previous
values of the derivative. Each successive integration step is thereby
reduced to a fixed weighted average of some number of the previous
(and for AM methods, one future/predicted) derivative data points.
For fixed step size, the fit and integration can be done ahead of
time, once, and apply to all mesh locations, for all variables to
be integrated and all integration steps.

The fixed-grid AB and AM methods can be combined, but then the error
estimate is a passive report useful only in ex post facto analysis.
The adaptive-grid methods we present here use the error estimate dynamically,
to adjust the step size. They still allow a set of weighting factors
to be used in all mesh locations and for all variables. However, the
weights must be recalculated with each new integration step. Happily,
the computational overhead demanded by our methods is negligible compared
to derivative evaluation for complex problems, especially as it does
not scale with the size of the mesh or number of variables to be integrated.

The AB and AM methods having been derived in the nineteenth century
\citep{BA}, their fixed weighting was customarily used to reduce
the computational overhead of each step. In a twenty-first century
context, however, maintaining fixed weights (and thus a fixed step
size) amounts to assuming that the derivatives are extremely computationally
inexpensive. This assumption is massively violated for nearly all
simulations done today; for instance, in radiation transport, unless
potentially hazardous simplifying assumptions are made, the radiation
field is a function of seven independent variables $\left(x,y,z,\theta,\phi,\nu,t\right)$,
i.e., spatial location, two angles to specify direction, frequency,
and time. Since the first six dimensions must be solved for every
derivative evaluation, this dominates the computational time required
for a radiation transport-hydrodynamics simulation.

Because of the working assumption of cheap derivatives and the handicap
of fixed step size, we believe the AB and AM methods have been neglected
by the simulation community in favor of methods such as Runge-Kutta
(RK) \citep{Runge}. The RK methods can be used in an implicit scheme
and allow an adaptive step size, at the unfortunate price of even
more derivative evaluations per integration step, increasing with
the order of convergence of the particular method selected. In the
case of the most commonly used RK method, the explicit fourth order
Runge-Kutta, this means derivatives must be evaluated four times for
each integration step.

The adaptive mesh ABM methods presented here embody a highly advantageous
combination of features. These methods require only two derivative
evaluations per integration step (independent of convergence order!),
are implicit and therefore avoid Courant-Friedrichs-Lewy conditioning
\citep{Courant}, provide arbitrarily high orders of convergence,
and use the error estimate afforded by combining the AB and AM methods
to maintain an approximately constant fractional error through each
integration step. The last three features in concert allow the size
of the integration step to be, potentially, extremely large. Two derivative
evaluations per integration step is the minimum possible for an implicit
scheme, as implicity requires knowledge of at least the derivative
at the present value of the independent variable and at the next data
point for which the method is implicitly solving.

The price extracted for the speed-ups just mentioned are that previous
values of the derivatives must remain accessible to the integration
routine. Codes that use single, rather than multi-step integration
methods methods may write their meshes to file (to be archived and
never touched by the code again) and then alter the mesh in place.
These codes would have to be rewritten in order to maintain code access
to roughly as many integration steps as the desired order of convergence.
We believe any penalty in either memory requirements or hard disk
lag will be far outweighed by the speed-ups the ABM methods make achievable
for problems in which derivative evaluation is very expensive.

Another pitfall to be avoided is Runge's phenomenon \citep{Runge Phenom},
where the error involved in an interpolation actually increases as
the order of the interpolating polynomial becomes very large, due
to a {}``ringing'' effect. The order of convergence can be tuned
for a particular problem in order to minimize Runge's phenomenon.
We demonstrate this tuning in Section \ref{sub:Parameter-Hunts},
where we find the order of convergence that requires the fewest integration
steps for a particular prescribed precision.

We explore the practical application of our ABM methods with a trial
problem of determining the Tolman-Oppenheimer-Volkoff (TOV) limit
\citep{Tolman} for the Oppenheimer-Volkoff (OV) equation of state
\citep{OV}.

\section{THEORY\label{sec:Theory}}

\subsection{Adams-Bashforth\label{sub:Adams-Bashforth}}

Our methods represent a discretization of the quantity we wish to
integrate numerically. Let us call the function we would like to calculate
and its derivative, respectively, $y\left(x\right)$ and $y'\left(x\right)$.
In complete generality we can quantize it thus:

\begin{equation}
y_{i+1}\equiv y\left(x_{i+1}\right)=y\left(x_{i}\right)+\int_{x_{i}}^{x_{i+1}}y'\left(x,y\right)\,\mathrm{d}x\label{eq:QuantizationOfY}\end{equation}

We shall quantize the derivative and other quantities with Latin letter
subscripts analogously, and call the initial conditions $y_{0}\equiv y(x_{0})$.
Note that multiple variables, coupled or not, can be integrated by
giving $y$ and $y'$ vector subscripts, as well. This subscript can
denote either a true, traditional vector, or simply a list of variables
all needing integration (e.g., $m$ and $P$, as will be discussed
in Section \ref{sec:TOV-Test-Problem}). For simplicity, we omit any
such subscript in the following derivation.

The Adams-Bashforth methods are based on a Lagrange polynomial approximation
to the derivative. Our method requires a new Lagrange polynomial at
each integration step $i$, which we shall call $P_{\mathrm{AB}\, i}$.
For an order of convergence $N$, we have:

\begin{equation}
P_{\mathrm{AB}\, i}\left(\bar{x}\right)\equiv\sum_{j=i-N+1}^{i}y'_{j}\prod_{k=i-N+1,k\neq j}^{i}\frac{\left(\bar{x}-x_{k}\right)}{\left(x_{j}-x_{k}\right)}\end{equation}

where $y'_{j}=y'\left(x_{j},y_{j}\right)$. Choosing $\bar{x}=x-x_{i}$
allows us to identify the origin according to $P_{\mathrm{AB}\, i}$
with the most recent integration step. Our analysis is simplified
if we make the following definition:

\begin{equation}
\psi_{\mathrm{AB}\, ij}\left(\bar{x}\right)\equiv\prod_{k=i-N+1,k\neq j}^{i}\frac{\left(\bar{x}-x_{k}\right)}{\left(x_{j}-x_{k}\right)}\end{equation}

Thus:

\begin{equation}
P_{\mathrm{AB}\, i}\left(\bar{x}\right)=\sum_{j=i-N+1}^{i}y'_{j}\psi_{\mathrm{AB}\, ij}\left(\bar{x}\right)\end{equation}

With this definition in hand, we can now derive our Adams-Bashforth
approximation to $y_{i+1}$. With our earlier selection of the origin
of $P_{\mathrm{AB}\, i}$, the integral in Equation \ref{eq:QuantizationOfY}
becomes:

\begin{eqnarray}
y_{\mathrm{AB}\, i+1} & \approx & y_{\mathrm{AB}}\left(x_{i}\right)+\int_{0}^{\Delta x_{i}}P_{\mathrm{AB}\, i}\left(\bar{x}\right)\,\mathrm{d}\bar{x}\nonumber \\
 & = & y_{\mathrm{AB}}\left(x_{i}\right)+\int_{0}^{\Delta x_{i}}\sum_{j=i-N+1}^{i}y'_{j}\psi_{\mathrm{AB}\, ij}\left(\bar{x}\right)\,\mathrm{d}\bar{x}\label{eq:Final Expression for yAB i+1}\\
 & = & y_{\mathrm{AB}}\left(x_{i}\right)+\sum_{j=i-N+1}^{i}\left(\int_{0}^{\Delta x_{i}}\psi_{\mathrm{AB}\, ij}\left(\bar{x}\right)\,\mathrm{d}\bar{x}\right)y'_{j}\nonumber \end{eqnarray}

Note that $\psi_{\mathrm{AB}\, ij}\left(\bar{x}\right)$ is a calculable
polynomial and can be integrated analytically. We implement that analytic
integration with the following definitions:

\begin{equation}
\Phi_{\mathrm{AB}\, i}\left(\bar{x}\right)\equiv\prod_{k=i-N+1}^{i}\left(\bar{x}-x_{k}\right)\label{eq:Phi def}\end{equation}

and

\begin{equation}
\phi_{\mathrm{AB}\, ij}\left(\bar{x}\right)\equiv\prod_{k=i-N+1,k\neq j}^{i}\left(\bar{x}-x_{k}\right)=\frac{\Phi_{\mathrm{AB}\, i}\left(\overline{x}\right)}{\left(\overline{x}-x_{j}\right)}\label{eq:phi def}\end{equation}

We use an equals sign loosely in Equation \ref{eq:phi def} because
the rightmost expression is undefined for $\overline{x}=x_{j}$ while
the middle expression has no such singularity. This is immaterial
to us, however, as we will evaluate the fraction analytically before
substituting in any specific values for variables. Using an equals
sign in a similar way, we can therefore express $\psi_{\mathrm{AB}\, ij}\left(\bar{x}\right)$
as:

\begin{equation}
\psi_{\mathrm{AB}\, ij}\left(\overline{x}\right)=\frac{\phi_{\mathrm{AB}\, ij}\left(\overline{x}\right)}{\phi_{\mathrm{AB}\, ij}\left(x_{j}\right)}\end{equation}

If we compute the product $\Phi_{\mathrm{AB}\, i}\left(\bar{x}\right)$
first and save the penultimate intermediate result, we need calculate
only the product of $N$ binomials and carry out $N-1$ synthetic
polynomial divisions of that product by binomials for each integration
step.

\subsection{Adams-Moulton\label{sub:Adams-Moulton}}

The derivation of the AM methods proceeds almost identically to the
AB methods, so we present them with little comment. Note that the
upper boundary of the sums and products in these equations is $i+1$,
in contrast to the $i$ limit of the AB method formulae.

\begin{equation}
P_{\mathrm{AM}\, i}\left(\bar{x}\right)\equiv\sum_{j=i-N+1}^{i+1}y'_{j}\prod_{k=i-N+1,k\neq j}^{i+1}\frac{\left(\bar{x}-x_{k}\right)}{\left(x_{j}-x_{k}\right)}\end{equation}

The key to linking the AB and AM methods into an ABM predictor-corrector
tandem is to insert the AB prediction for $y_{i+1}$ into the calculation
of $y'_{i+1}$. That is,

\begin{equation}
y'_{i+1}\equiv y'\left(x_{i+1},y_{\mathrm{AB}\, i+1}\right)\label{eq:Insert yAB i+1 into y'i+1}\end{equation}

With this identification, the derivation proceeds as before.

\begin{equation}
\psi_{\mathrm{AM}\, ij}\left(\bar{x}\right)\equiv\prod_{k=i-N+1,k\neq j}^{i+1}\frac{\left(\bar{x}-x_{k}\right)}{\left(x_{j}-x_{k}\right)}\end{equation}

\begin{equation}
P_{\mathrm{AM}\, i}\left(\bar{x}\right)\equiv\sum_{j=i-N+1}^{i+1}y'_{j}\psi_{\mathrm{AM}\, ij}\left(\bar{x}\right)\end{equation}

\begin{eqnarray}
y_{\mathrm{AM}}\left(x_{i+1}\right) & \approx & y_{\mathrm{AM}}\left(x_{i}\right)+\int_{0}^{\Delta x_{i}}P_{\mathrm{AM}\, i}\left(\bar{x}\right)\,\mathrm{d}\bar{x}\nonumber \\
 & = & y_{\mathrm{AM}}\left(x_{i}\right)+\int_{0}^{\Delta x_{i}}\sum_{j=i-N+1}^{i+1}y'_{j}\psi_{\mathrm{AM}\, ij}\left(\bar{x}\right)\,\mathrm{d}\bar{x}\label{eq:Final yAM xi+1}\\
 & = & y_{\mathrm{AM}}\left(x_{i}\right)+\sum_{j=i-N+1}^{i+1}\left(\int_{0}^{\Delta x_{i}}\psi_{\mathrm{AM}\, ij}\left(\bar{x}\right)\,\mathrm{d}\bar{x}\right)y'_{j}\nonumber \end{eqnarray}

We implement the calculation in a similar fashion as well.

\begin{equation}
\Phi_{\mathrm{AM}\, i}\left(\bar{x}\right)\equiv\prod_{k=i-N+1}^{i+1}\left(\bar{x}-x_{k}\right)\end{equation}

\begin{equation}
\phi_{\mathrm{AM}\, ij}\left(\bar{x}\right)\equiv\prod_{k=i-N+1,k\neq j}^{i+1}\left(\bar{x}-x_{k}\right)=\frac{\Phi_{\mathrm{AM}\, i}\left(\overline{x}\right)}{\left(\overline{x}-x_{j}\right)}\end{equation}

\begin{equation}
\psi_{\mathrm{AM}\, ij}\left(\overline{x}\right)=\frac{\phi_{\mathrm{AM}\, ij}\left(\overline{x}\right)}{\phi_{\mathrm{AM}\, ij}\left(x_{j}\right)}\end{equation}

Ordinarily, the AB and AM methods require some other method, such
as RK, initially to supply enough derivative points to allow the use
of the desired order of method. These methods are designed to maintain
a particular precision, so instead we have implemented a boot-strapping
technique, where $y_{0}$ is supplied as the boundary condition, $y_{1}$
is calculated using a first-order AB and second-order AM method, $y_{2}$
is calculated using a second- and third-order tandem, etc., until
enough data points have been calculated that we can proceed with the
desired method. This can be elegantly implemented by replacing the
lower limits of sums and products with zero when they would be negative.
We will explore the consequences of this technique in Section \ref{sec:Polynomial-Test-Problem} 

For a production-quality data run, the step sizes need only be chosen
small enough that the desired precision is maintained throughout the
bootstrapping phase.

\subsection{Selecting A Step Size}

For most purposes, it is the precision of the final answer, not the
specific number of steps or specific step size that is of interest.
The power of an explicit-implicit predictor-corrector method is that
it allows the integrator to maintain an approximately constant relative
size of correction term throughout an integration. The user of the
technique we present, therefore, can specify the desired precision
of integration directly. In this section, we demonstrate how.

The two phases of a tandem ABM method of orders $N$ and $N+1$ estimate
the quantity to be integrated with orders of convergence that differ
by one. Subtracting the two estimates therefore gives a correction
term of order $N+1$ in the step size. Dividing by the prediction
gives the fractional or relative correction:

\begin{equation}
\epsilon_{i}=\frac{y_{\mathrm{AM}\, i+1}-y_{\mathrm{AB}\, i+1}}{y_{\mathrm{AB}\, i+1}}\propto O\left(\Delta x_{i}^{N+1}\right)\label{eq:Fractional Error}\end{equation}

Let us call the constant of proportionality in the above relation
$k_{i}$. We then find that:

\begin{equation}
\frac{\epsilon_{i+1}}{\epsilon_{i}}=\frac{k_{i+1}}{k_{i}}\left(\frac{\Delta x_{i+1}}{\Delta x_{i}}\right)^{N+1}\label{eq:Ratio of errors}\end{equation}

If we desire an ordained fractional correction of $E$ in each integration
step, let us simply set $\epsilon_{i+1}=E$ in Equation \ref{eq:Ratio of errors}.
We thereby derive a prescription for the next step size:

\begin{equation}
\Delta x_{i+1}=\left(\sqrt[N+1]{\frac{Ek_{i}}{\epsilon_{i}k_{i+1}}}\right)\Delta x_{i}\end{equation}

Without additional evaluations of the derivative, we do not have any
means of estimating the ratio $k_{i}/k_{i+1}$. Evaluating additional
derivatives would, of course, defeat the purpose of the methods we
are presenting, so instead let us observe that the ratio enters into
the prescription for the next step size only inside a root that in
practice may be as high as 9$^{\text{th}}$ or 10$^{\text{th}}$,
as we will see in Section \ref{sec:TOV-Test-Problem}. Assuming that
the ratio is unity will typically result in only small changes in
$\Delta x$ from one step to the next. Furthermore, adjusting the
step size with every integration step allows the method to continue
adapting to a transiently large ratio over a few integration steps.
In addition, it may be possible for the derivative evaluation procedures
themselves to detect regimes where the ratio is likely to deviate
more from unity, such as near phase changes, and to signal the step
size update procedure to shrink the step size in caution, though this
added sophistication is beyond the scope of this work. Assuming that
the ratio is unity, then, leaves us with the final prescription:

\begin{equation}
\Delta x_{i+1}=\left(\sqrt[N+1]{\frac{E}{\epsilon_{i}}}\right)\Delta x_{i}\label{eq:final step size}\end{equation}

When $y$, etc., are vectors, we use the largest component of $\epsilon_{i}$
in Equation \ref{eq:final step size}. We demonstrate that that equation's
step size prescription maintains the expected order of convergence
and error tolerances with an artificial test problem in Section \ref{sec:Polynomial-Test-Problem}
and in a real-world integration problem in Section \ref{sec:TOV-Test-Problem}.

\section{Polynomial Test Problem\label{sec:Polynomial-Test-Problem}}

In order to verify that our implementation of the ABM methods has
the desired orders of convergence, we first devised a simple polynomial
test problem:

\begin{equation}
y'\left(x\right)=\left(x-1\right)\left(x-2\right)\left(x-3\right)\left(x-4\right)=x^{4}-10x^{3}+35x^{2}-50x+24\label{eq:Polynomial Test Problem}\end{equation}

with the boundary condition

\begin{equation}
y\left(0.5\right)=1\label{eq:Polynomial Test Problem boundary}\end{equation}

\begin{figure}[H]
\includegraphics{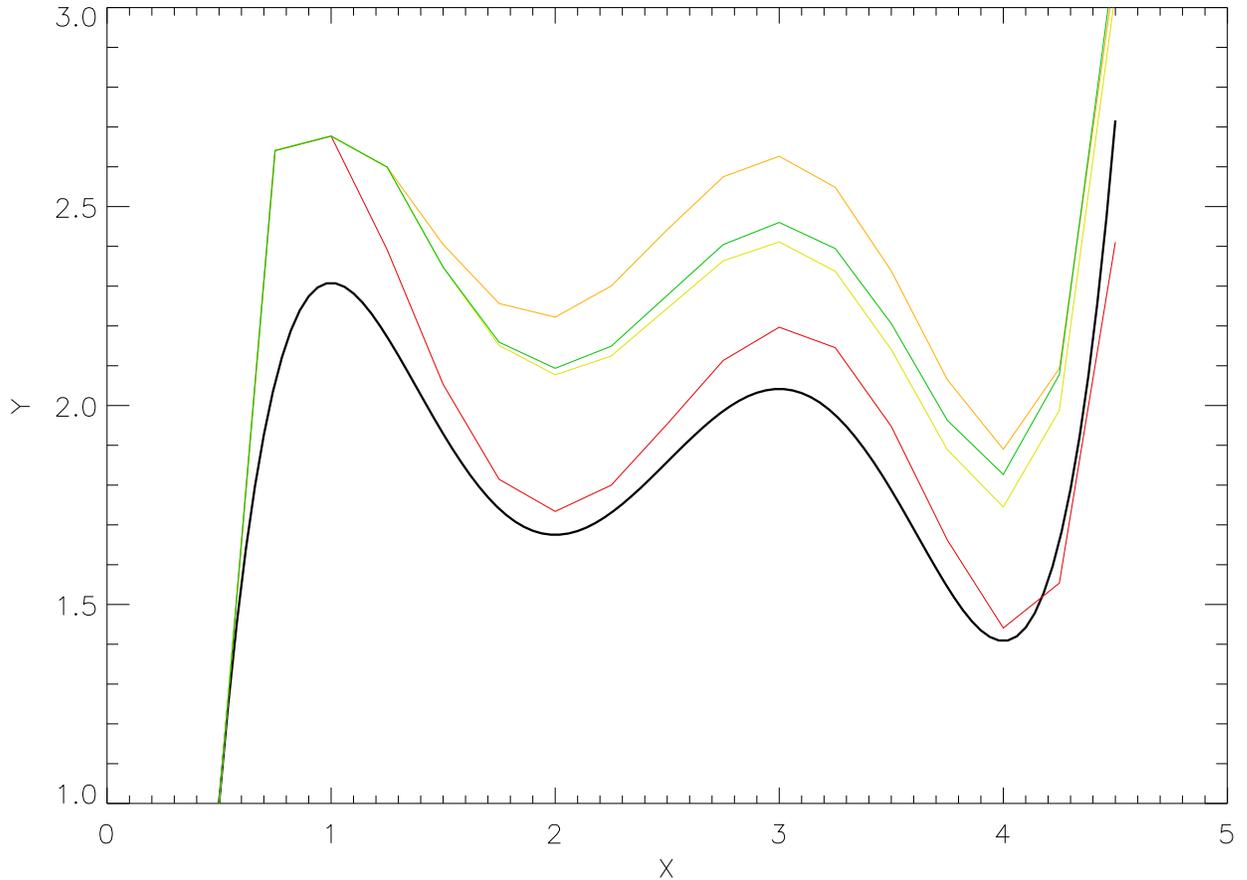}\caption{\emph{Adams-Bashforth solutions to polynomial test problem} The analytic
solution to the test problem $y'\left(x\right)=\left(x-1\right)\left(x-2\right)\left(x-3\right)\left(x-4\right)$;
$y\left(0.5\right)=1$ is plotted in bold black. Numerical solutions
using a fixed grid spacing of 0.25 and AB methods of order 1 (red),
2 (orange), 3 (yellow) and 4 (green) are plotted as well. Observe
that each successively higher order method deviates from the common
trajectory one integration step later. This is a result of the boot-strapping
technique we employ whereby the first few data points are generated
by successively higher order methods until the desired order of convergence
is reached. In this simple fixed-grid test case, boot-strapping produces
unacceptably large errors in the first few integration steps. In production
runs employing an adaptive mesh, however, the first integration step
size is specified to be small enough that even a first-second order
AB-AM predictor-corrector combined method produces errors smaller
than the desired tolerance. Our implementation will then scale up
the step size appropriately as the integration proceeds.\label{fig:AB poly solutions}}
\end{figure}

In Figure \ref{fig:AB poly solutions} we have plotted the analytic
solution $\left(y=x^{5}/5-5x^{4}/2+35x^{3}/3-25x^{2}+24x-727/120\right)$,
along with solutions calculated by Adams-Bashforth-only (no correction)
methods of order 1 (red), 2 (orange), 3 (yellow) and 4 (green). Without
the correction of an AM method to guide us, we maintain a fixed grid.
We have chosen the relatively large step size of $\Delta x=0.25$
in order to produce noticeable errors. We see that each consecutively
higher order method deviates from the common trajectory one step later-
i.e., all methods give the same result for the first two integration
steps, all but the first order method give the same result for the
next step, all but the first and second order methods give the same
result for the step after that, etc. This demonstrates that we have
implemented the bootstrapping technique discussed in Section \ref{sub:Adams-Moulton}
correctly.

\begin{figure}[H]
\includegraphics{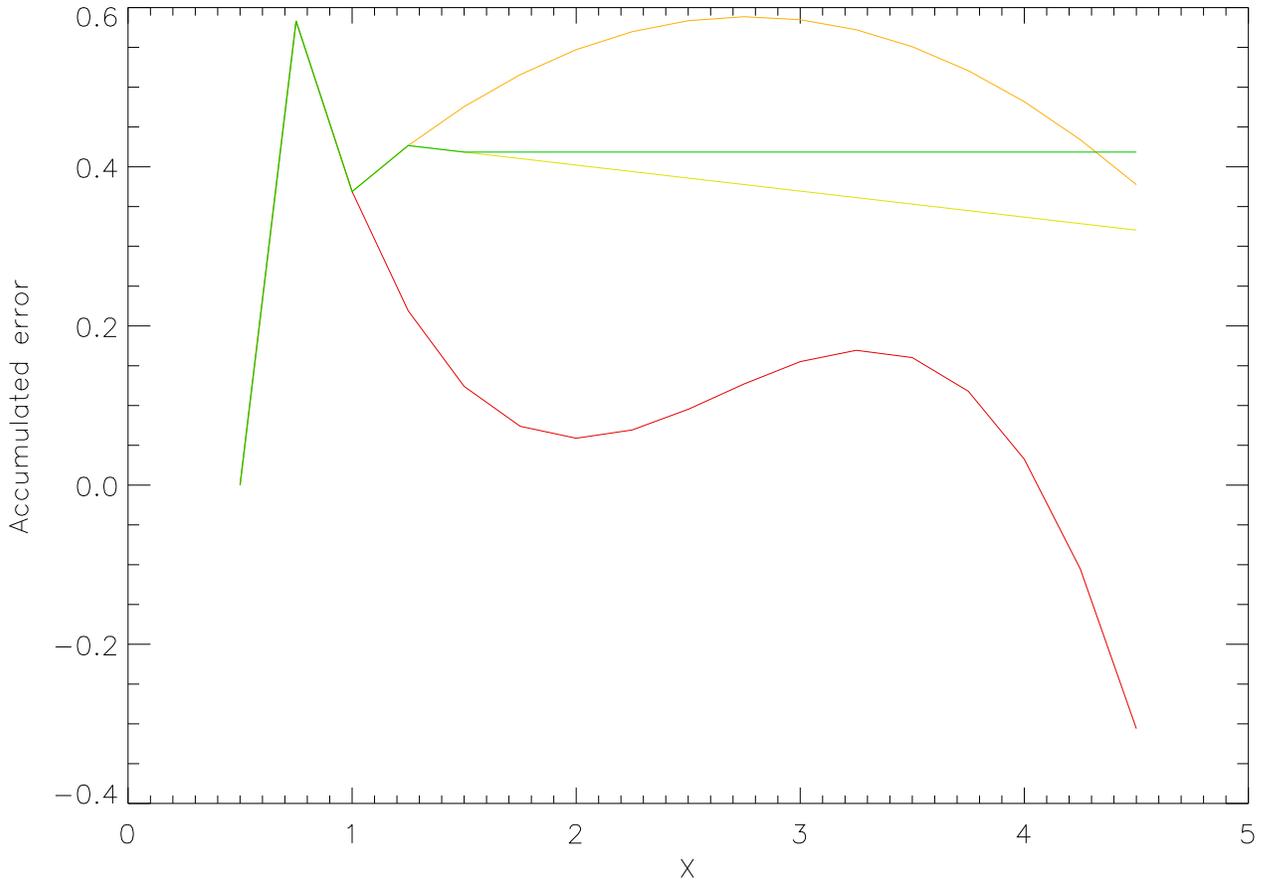}\caption{\emph{Accumulated error from Adams-Bashforth solutions to polynomial
test problem} We plot the accumulated errors with the same color convention
as Figure \ref{fig:AB poly solutions}. Note the polynomial behavior
of the accumulated error of each method after its initial boot-strapping
phase. The first order method's error is cubic, the second order method's
error is parabolic, third order's error is linear, and, most importantly,
the fourth order AB method accumulates no error greater than roundoff.
This proves that our AB methods demonstrate the expected order of
convergence. \label{fig:AB poly Accumulated Error}}
\end{figure}

In Figure \ref{fig:AB poly Accumulated Error} we plot the running
error accumulated by each of the AB methods previously plotted in
Figure \ref{fig:AB poly solutions}. Here we see that the accumulated
errors of each consecutive method exhibit polynomial behavior of decreasing
order, culminating in a fourth order AB method that accumulates no
error greater than double-precision roundoff for our fourth-order
polynomial derivative. This proves that our implementation of AB methods
exhibits the expected orders of convergence.

\begin{figure}[H]
\includegraphics{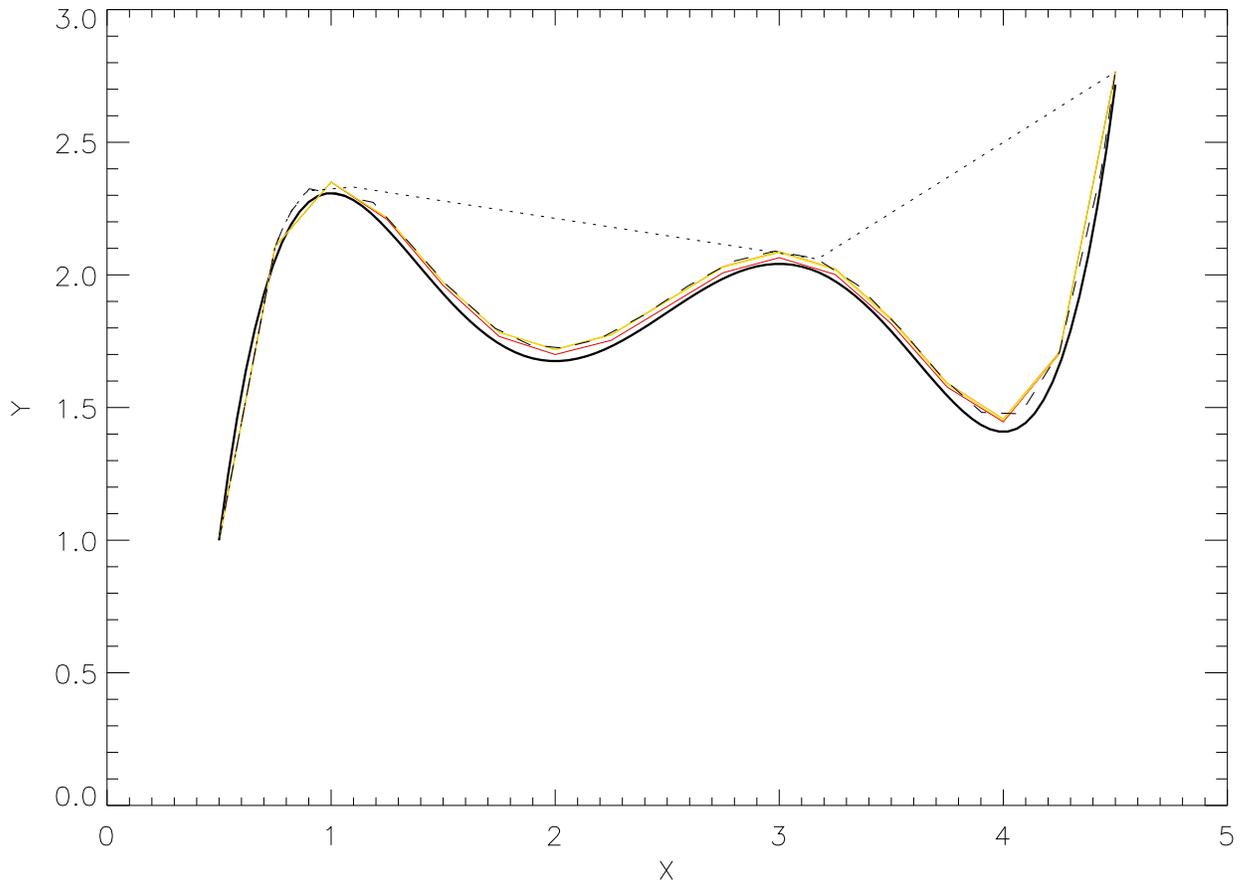}

\caption{\emph{Adams-Bashforth-Moulton solutions to polynomial test problem}
We plot various solutions to our polynomial test problem. As before,
the analytic solution is bold black. Fixed-grid (step size 0.25, as
in Figure \ref{fig:AB poly solutions}) ABM solutions are plotted
in color. Each solution is plotted in the same color as the AB-only
solution in Figure \ref{fig:AB poly solutions} of the same AB order.
Adding the AM correction then increases the order of convergence by
1. We also plot two additional solutions on this figure. The dashed
curve is the order 3-4, adaptive grid ABM method. The dotted curve
is the order 4-5, adaptive grid ABM method. Note that the adaptive-grid
3-4 ABM solution must follow the curve, maintaining small step sizes.
After the initial boot-strapping, the adaptive-grid 4-5 ABM method
arrives at its final result in only two integration steps.\label{fig:ABM poly solutions}}
\end{figure}

In Figure \ref{fig:ABM poly solutions} we plot the solutions to our
polynomial test problem achieved by ABM methods of various degrees.
The fixed-grid ABM methods are plotted in the same color as the (fixed-grid)
AB-only solution of the same AB order in Figure \ref{fig:AB poly solutions}.
For instance, the AB method of order 2 and the ABM method of order
2-3 are both plotted in orange in Figures \ref{fig:AB poly solutions}
and \ref{fig:ABM poly solutions}, respectively. We omit the fixed-grid
ABM method of order 4-5 because it is coincident with the fixed-grid
ABM method of order 3-4. We also plot two adaptive-grid methods in
Figure \ref{fig:ABM poly solutions}. The adaptive-grid ABM method
of order 3-4 is plotted as a dashed curve, while the adaptive-grid
ABM method of order 4-5 is plotted as a dotted curve. Even with the
artificially large integration steps (or first integration step, in
the adaptive-grid cases), we see that adding Adams-Moulton correction
to the Adams-Bashforth prediction results in much smaller errors. 

\begin{figure}[H]
\includegraphics{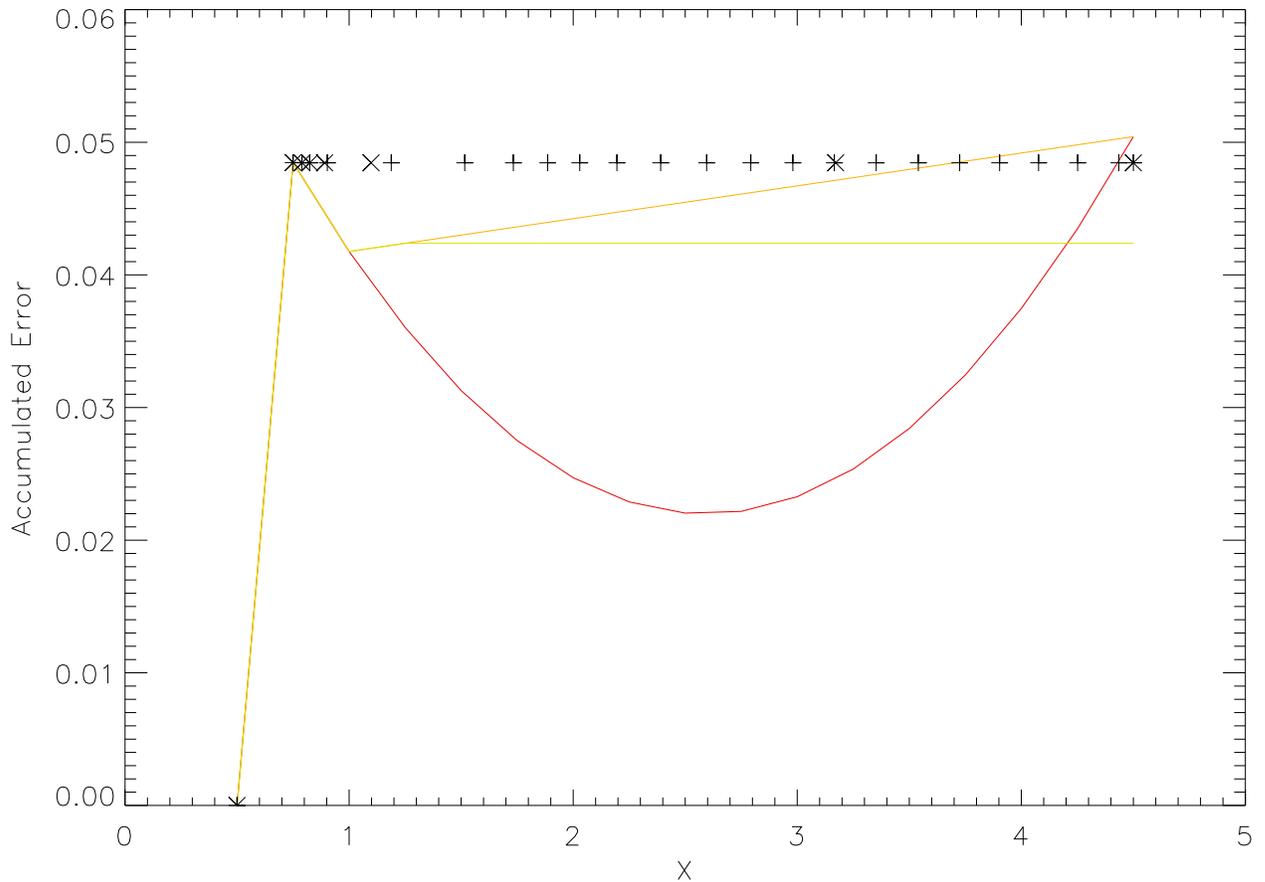}\caption{\emph{Accumulated error from Adams-Bashforth-Moulton solutions to
Polynomial test problem} We plot accumulated error for all numeric
methods described in Figure \ref{fig:ABM poly solutions}. Colored
curves denote the errors accumulated by curves of the corresponding
color in that figure. For clarity, accumulated errors of the 3-4 adaptive
grid ABM and 4-5 adaptive grid ABM methods are now plotted as +'s
and X's, respectively. Note that each fixed-grid ABM curve displays
an order of convergence greater by one than its corresponding AB-only
method, as expected. This proves that the fixed-grid ABM methods demonstrate
the expected order of convergence. Also note that, for example, the
method with parabolic accumulated error (ABM order 1-2) ends its bootstrapping
phase one integration step earlier than the parabolic-error AB-only
method (AB Order 2). This is also expected behavior for, and a benefit
of, an implicit corrector.\label{fig:ABM poly Accumulated Error}}
\end{figure}

In Figure \ref{fig:ABM poly Accumulated Error} we plot the error
accumulated by the ABM methods discussed in Figure \ref{fig:ABM poly solutions}.
We see that the fixed-grid methods display similar decreasing-order
polynomial behavior with increasing order of convergence. As befits
methods with a correction phase that increases the order of convergence
by one, each curve here is a lower-by-one order polynomial compared
to the curve of accumulated error of the equal AB order method's solution
(and same color) plotted in Figure \ref{fig:AB poly Accumulated Error}.
For instance, the red curves in \ref{fig:AB poly Accumulated Error}
and \ref{fig:ABM poly Accumulated Error} refer to the error accumulated
by the AB method of order 1 and the ABM method of order 1-2, respectively,
but the former displays cubic behavior while the latter displays parabolic.
In addition, the parabolic-error curve in Figure \ref{fig:ABM poly Accumulated Error}
(red) deviates from the common trajectory one integration step sooner
than the parabolic-error curve in Figure \ref{fig:AB poly Accumulated Error}
(orange). These patterns confirm that our fixed-grid ABM methods display
the expected $N$+1 order of convergence.%

We also plot the error accumulated by the adaptive-grid ABM methods
of order 3-4 and 4-5, this time as $+$'s and $\times$'s, respectively,
for clarity. We see that both have negligible accumulated error after
the initial (artificially large) integration step. In this artificial
problem, there are no error terms after the correction phase of the
adaptive-grid ABM method of order 3-4, so we expected no error to
accumulate. However, the correction phase is necessary in each integration
step to capture all the behavior of the derivative, and the correction
remains roughly constant in magnitude with each step. The step size
thus remains roughly constant as well, because this method has no
way of determining that it could increase its step size without increasing
its error.

\begin{figure}[H]
\includegraphics{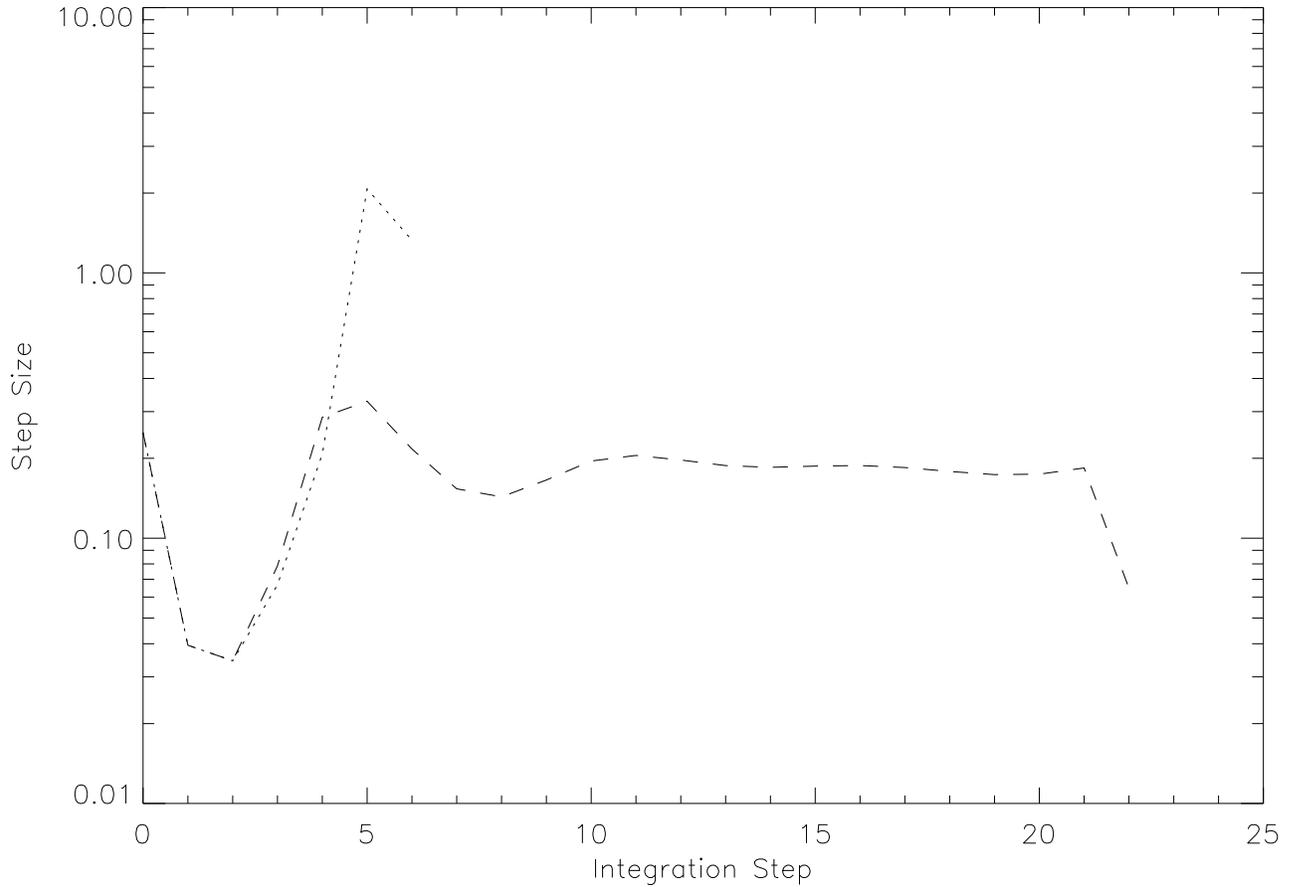}\caption{\emph{Step size of adaptive-mesh Adams-Bashforth-Moulton solutions
to polynomial test problem} We plot the step sizes of the Order 3-4
and Order 4-5 ABM methods as, respectively, dashed and dotted curves,
similar to Figure \ref{fig:ABM poly solutions}. For our fourth-order
polynomial test problem, there are, in fact, no further error terms
after the Order 3-4 method's AM correction phase, but there is no
way for the method to determine that. It must therefore maintain a
relatively constant step size. The Order 4-5 ABM solution, in contrast,
captures all information about the derivative in the AB phase already,
and the AM phase confirms that there is no correction to be made.
The step size is therefore increased dramatically. The divergent behavior
of the two methods, with the precondition that the fixed-grid methods
are known to converge as expected (See Figure \ref{fig:ABM poly Accumulated Error}),
demonstrates that the ABM methods maintain their expected order of
convergence even when the step size is allowed to change.\label{fig:ABM poly dx vs i}}
\end{figure}

The adaptive-grid ABM method of order 4-5, in contrast, captures the
entire behavior of the derivative in the AB phase, with the AM correction
confirming that the errors are no larger than roundoff. After its
initial boot-strapping phase, the adaptive-grid ABM method of order
4-5 can then increase its step size geometrically. Although we have
no theoretical bound on how large the step size could be made, for
caution in case of the coincidental vanishing of just one error term,
in practice we limit its growth to geometric, with a ratio of 3. The
adaptive-grid ABM method of order 4-5 arrives at its final answer
in only two integration steps after its initial boot-strapping. We
plot the step size for these two methods in Figure \ref{fig:ABM poly dx vs i}.
The Order 3-4 and 4-5 methods are plotted as dashed and dotted, respectively,
analogously to Figure \ref{fig:ABM poly solutions}.

The errors in the initial integration step for all the methods discussed
thus far have been quite large, but this is only because we have deliberately
selected a very large step size parameter in order to highlight the
error behavior of all these methods. In a production-quality data
run using adaptive-grid ABM methods, the initial step size parameter
would be selected so as to give negligible error in the first integration
step, and thereafter would be rapidly increased to a size commensurate
with the desired precision of integration. The step sizes decrease
markedly in both of the methods shown in Figure \ref{fig:ABM poly dx vs i}
in the last integration step because the integrations have already
reached the upper limit.

\section{TOLMAN-OPPENHEIMER-VOLKOFF TEST PROBLEM\label{sec:TOV-Test-Problem}}

\subsection{Motivation and Introduction\label{sub:Motivation-and-Introduction}}

Since the ABM methods passed our polynomial test problem, we devised
a more complex problem to test our implementation in a real-world
setting. We selected the famous problem of deriving the maximum possible
mass (as measured by an observer at infinity) of material that can
be supported hydrostatically against gravitational collapse- in short,
the most massive possible neutron star. \citet{OV} (OV) discovered
the existence of this limit for any equation of state obeying relativistic
causality. They also derived the numerical value of 0.71 M$_{\odot}$
for the limit under the assumption that the only contribution to pressure
is that of neutron momentum arising from the Pauli exclusion principle.
Their value of the radius corresponding to this mass is 9.5 km. For
this problem, we expand the $y$ variable from Sections \ref{sec:Theory}
and \ref{sec:Polynomial-Test-Problem} into a vector whose elements
are pressure and mass-energy enclosed within the radius.We discuss
the details of this problem throughout the rest of this section.

\subsection{Hydrostatic Equilibrium\label{sub:Hydrostatic-Equilibrium}}

The problem of finding the TOV limit assumes bodies in non-rotating,
spherical, hydrostatic equilibrium. The Schwarzschild metric \citep{Schwarzschild}
describes this geometry:

\begin{equation}
\mathrm{d}s^{2}=-e^{2\Phi}\mathrm{d}t^{2}+e^{2\Lambda}\mathrm{d}r^{2}+r^{2}\mathrm{d}\Omega^{2}\end{equation}

where $s$ is the spacetime interval, $t$ is the timelike variable,
$r$ is the radial spatial variable, $\Omega$ is the solid angle,
and $\Phi$ and $\Lambda$ are metric functions of $r$.

From there, the equations describing non-rotating spheres in hydrostatic
equilibrium can be easily derived (e.g., \citet{MTW} (MTW), p. 600,
their equations 23.19 and 23.22):

\begin{equation}
m(r)=\int_{0}^{r}4\pi\bar{r}^{2}\rho\,\mathrm{d}\bar{r}+m(0)\label{eq:m of r integral}\end{equation}

\begin{equation}
\frac{\mathrm{d}P}{\mathrm{d}r}=-\frac{\left(\rho+P\right)\left(m+4\pi r^{3}P\right)}{r\left(r-2m\right)}\label{eq:raw OV}\end{equation}

where $m$ is the mass enclosed within $r$, $\rho$ is the mass-energy
density expressed as an energy per volume, and $P$ is the pressure,
and all of these variables are functions of $r$. Reintroducing factors
of $G$ and $c$ omitted in MTW, taking the derivative with respect
to $r$ in Equation \ref{eq:m of r integral}, and explicitly stating
the variable dependencies gives

\begin{equation}
\frac{\mathrm{d}m(r)}{\mathrm{d}r}=\frac{4\pi}{c^{2}}r^{2}\rho(r)\label{eq:dmdr final}\end{equation}

\begin{equation}
\frac{\mathrm{d}P(r)}{\mathrm{d}r}=-\frac{G}{c^{2}r^{2}}\left(\rho(r)+P(r)\right)\left(m(r)+\frac{4\pi}{c^{2}}r^{3}P(r)\right)\left(1-\frac{2Gm(r)}{c^{2}r}\right)^{-1}\label{eq:dpdr final}\end{equation}

These equations by themselves do not form a closed system. We discuss
their closure in Section \ref{sub:Equation-of-State}.

\subsection{Equation of State\label{sub:Equation-of-State}}

In order to form a closed system of equations, we must supplement
Equations \ref{eq:dmdr final} and \ref{eq:dpdr final} with an equation
of state (EOS), i.e. a relationship between $P$ and $\rho$. In order
to compare to the OV results, we use the same EOS. The OV EOS assumes
that pressure results only from neutrons with the minimum possible
momenta allowed by the Pauli exclusion principle. \citealp{Kippenhahn Weigert}
(p. 118, hereafter KW) derive an analogous relation for electrons.
The identity of the particle giving rise to the pressure enters into
the derivation only in the mass, so to adapt their derivation we simply
replace the mass of an electron, $m_{e}$, with the mass of a neutron,
$m_{n}$. Following KW, and with $n$ the number density of neutrons:

\begin{equation}
P(n)=\frac{\pi m_{n}^{4}c^{5}}{3h^{3}}\left(x\left(2x^{2}-3\right)\sqrt{x^{2}+1}+3\ln\left(x+\sqrt{x^{2}+1}\right)\right)\label{eq:P(n)}\end{equation}

where

\[
x\equiv\frac{h}{2m_{n}c}\sqrt[3]{\frac{3n}{\pi}}\]

The energy density has contributions both from rest mass and from
the internal kinetic energy of the particles, which we shall call
$U$.

\begin{equation}
\rho(n)=m_{n}c^{2}n+U(n)\end{equation}

We again adapt a KW expression (p. 122), this time for $U$:

\begin{equation}
\rho(n)=m_{n}c^{2}n+\frac{\pi m_{n}^{4}c^{5}}{3h^{3}}\left(3x\left(2x^{2}+1\right)\sqrt{x^{2}+1}-8x^{3}-3\ln\left(x+\sqrt{x^{2}+1}\right)\right)\label{eq:rho(n)}\end{equation}

We will need to solve Equations \ref{eq:P(n)} and \ref{eq:rho(n)}
numerically to eliminate $n$.

\subsection{A Single Integration\label{sub:A-Single-Integration}}

We use an ABM method to integrate Equations \ref{eq:dmdr final} and
\ref{eq:dpdr final} over $r$, supplemented by Equations \ref{eq:P(n)}
and \ref{eq:rho(n)}, in the following manner. We select a value for
central pressure. This single boundary condition spans the solution
space for our system of equations. We then invert Equation \ref{eq:P(n)}
numerically in order to determine the number density of neutrons $n$.
The number density is inserted into Equation \ref{eq:rho(n)} to find
$\rho$. Since we have selected an EOS intended for use in neutron
stars, we select an initial integration step size, that is, $\Delta r_{0}$,
so small as to give negligible errors even in the first integration
step, which given our bootstrapping technique will be an order 1-2
ABM method. We have determined that 10 cm is quite sufficiently small.
We then have enough information to determine the derivatives of both
quantities to be integrated, $m(r)$ and $P(r)$, via Equations \ref{eq:dmdr final}
and \ref{eq:dpdr final}. We update the $\Delta r$ using the predictor-corrector
technique. The integration gives us the next value of $P(r)$, so
we can repeat this procedure, increasing the order of the ABM method
by one each step until the maximum desired order is reached. We halt
the integration when $P(r)$ reaches or overshoots zero. The total
mass and radius of the hydrostatic sphere, $M$ and $R$, respectively,
are defined as the final values of $m(r)$ and $r$.

\subsection{Parameter Hunts\label{sub:Parameter-Hunts}}

Some experimentation with different central pressures in the procedure
outlined in Section \ref{sub:A-Single-Integration} will quickly reveal
that the neutron star corresponding to the TOV limit in this EOS must
have a central pressure that lies between 10$^{\text{35}}$ and 10$^{\text{36}}$
erg cm$^{\text{-3}}$. We use a trinary sieve to reduce that range
to 3.631382 x 10$^{\text{35}}$ erg cm$^{\text{-3}}$ $\pm$ 16 in
the final two decimal places. During the sieving, we use an initial
step size of 10 cm, an ABM method of order 6-7, and a desired stepwise
tolerance of 10$^{\text{-8}}$. To ensure that the each integration
will terminate in a reasonable period of time even in the face of
the dramatic vanishing of pressure near the surface, we force the
step size to remain at least 10 cm. The midpoint of our pressure range,
with the same integration parameters except an ABM order of 10-11,
gives a total mass of 0.71017188 M$_{\odot}$ and a radius of 9.16233
km. \citet{OV} arrived at values of 0.71 M$_{\odot}$ and 9.5 km.
We have achieved perfect agreement in total mass within the precision
of their published results, but we do have a discrepancy of some 300
m or roughly 3.5\% in radius. OV do not disclose their method of numerical
integration, but we believe it is certainly plausible that a pre-WWII
numerical integration would have a precision of worse than 3.5\% in
the independent variable. This is especially true when its precise
value was then of so little interest compared to the mass, and when
the final radius has so little bearing on the former: the density
in the last few percent of radius is less by many orders of magnitude
than the central density, so the thickness of the crust is negligible
with respect to the total mass.

With superhigh-precision estimates in hand for the mass and radius,
we then conduct a parameter sweep in the tolerance and order of convergence
for what combinations produce the best combination of agreement with
the superhigh-precision results and few steps necessary to complete
the integration. We have compiled our results in Table \ref{tab:Integration-Steps-Required}.

\begin{table}[H]
\includegraphics{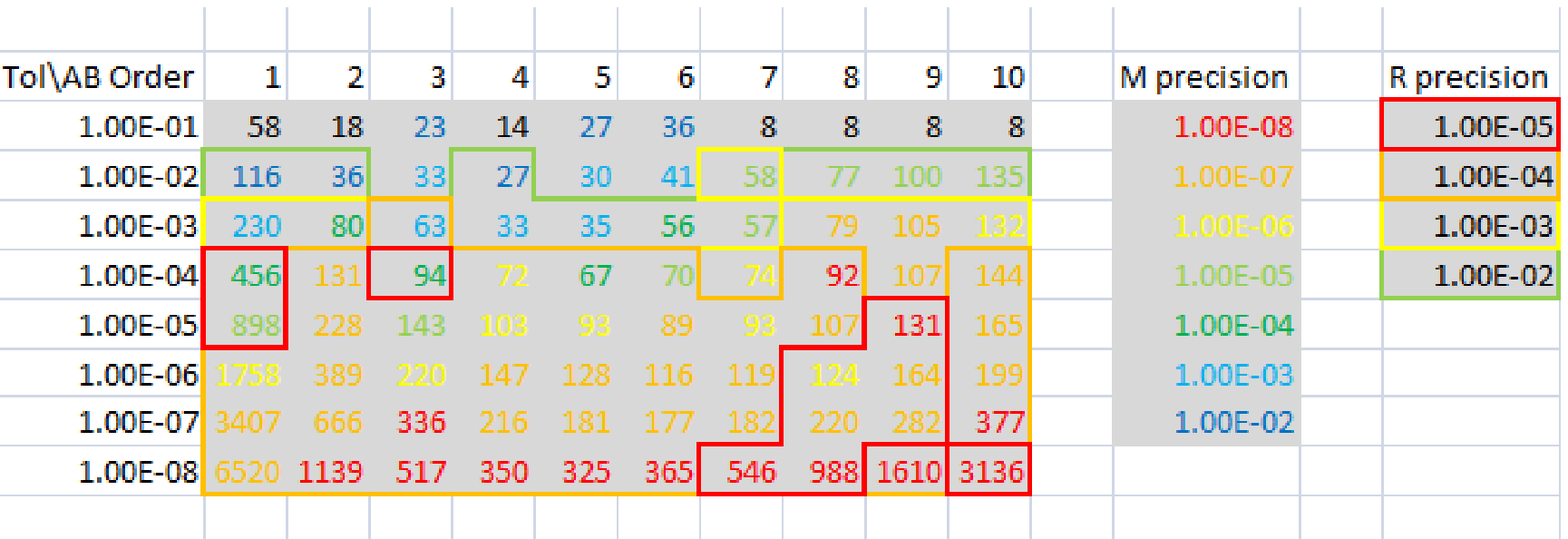}\caption{\emph{Integration Steps Required for Different Parameters, and the
Precision of Corresponding Data} We tabulate the integration steps
required to arrive at a full solution for various ABM adaptive-grid
methods (listed in the table by AB-phase order only) and for various
target step-wise maximum fractional AM corrections. The number listed
in the large table at left is the number of integration steps. The
precision with which a given set of parameters will match the total
mass as calculated by the superhigh-precision method in the lower
right corner is indicated by the color in which the integration step
is written. For instance, parameter sets that deliver solutions with
total neutron star mass matching the superhigh-precision parameters
within 1\% are written in dark blue. The precision with which a parameter
set matches the superhigh-precision radius are shown as colored contours.
For instance, parameter sets delivering a radius within one part in
1000 of the superhigh-precision radius are inside the yellow contour.
For maximal agreement with the superhigh-precision solution in total
M and total R, we find that an ABM order of 9-10 coupled with a desired
stepwise correction tolerance of 10$^{\text{-5}}$ arrives at a solution
in only 131 steps, while we can achieve a 1\% solution in 27 steps
with an ABM order of 4-5 coupled with a stepwise correction tolerance
of 10$^{\text{-2}}$. These two parameter sets are explored in more
detail throughout this section.\label{tab:Integration-Steps-Required}}
\end{table}

We list the number of steps required for each combination of parameters
to terminate in Table \ref{tab:Integration-Steps-Required} and higlight
the precision of agreement with the high-precision results for mass
and radius using colors. The color of the font denotes the precision
of agreement with the mass value of 0.71017188 M$_{\odot}$. Contour
lines denote the precision of agreement with the radius value of 9.16233
km. We find that two parameter combinations are especially efficient
for their desired precision. A tolerance of 10$^{\text{-2}}$ and
an ABM order of 4-5 achieves 1\% precision in both mass and radius
in only 27 integration steps. A tolerance of 10$^{\text{-5}}$ and
an ABM order of 9-10 achieves maximal precision in mass and radius
(10$^{\text{-8}}$ and 10$^{\text{-5}}$, respectively) in 131 integration
steps. We will refer to these combinations as the low precision and
high precision solutions, respectively, henceforth. The remainder
of this section is devoted to investigating these specific combinations
in greater detail.

\begin{figure}[H]
\includegraphics{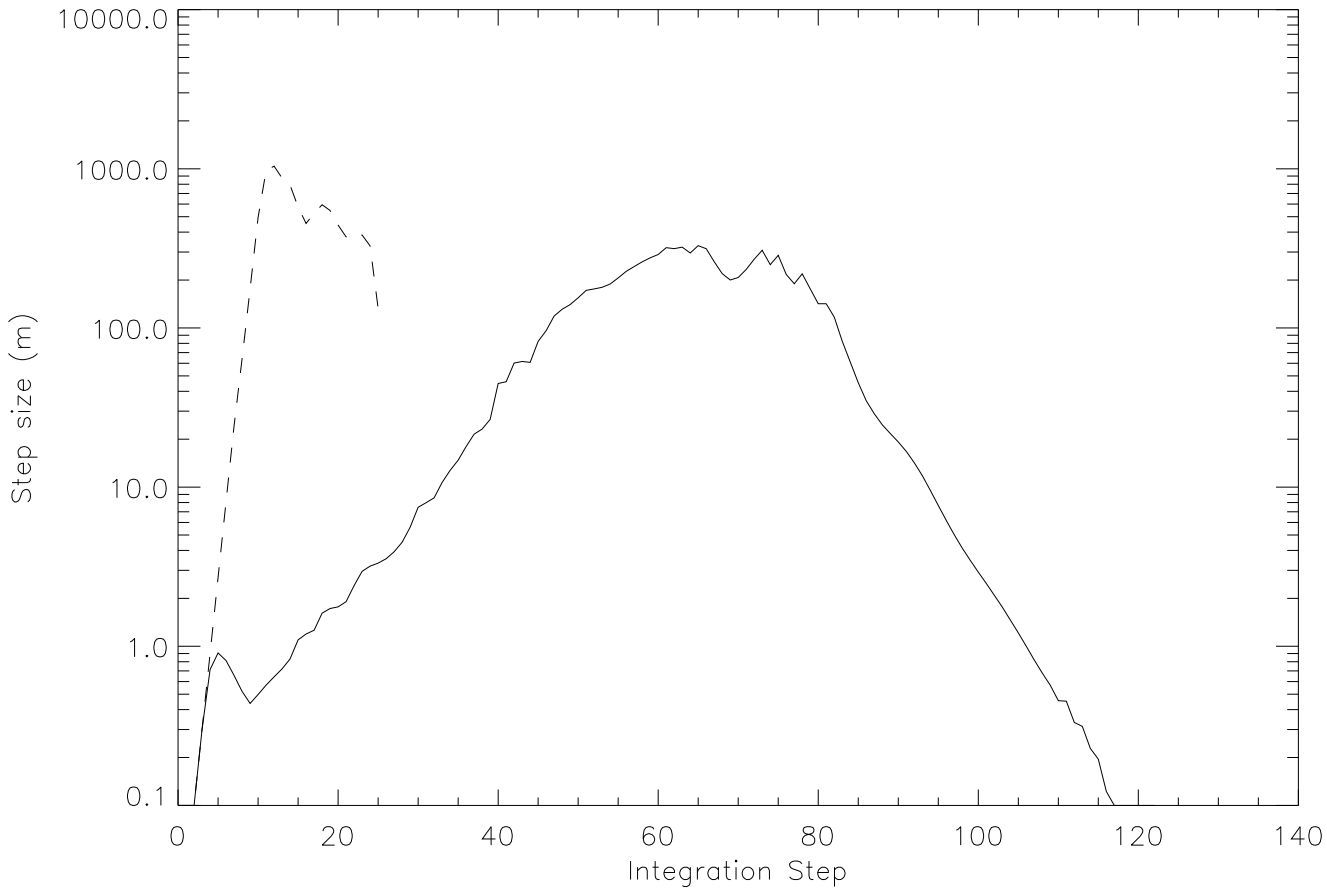}\caption{\emph{Step size for high (solid) and low (dashed) precision ABM solutions
to the Tolman-Oppenheimer-Volkoff test problem} We see that both solutions
exhibit initial exponential growth of step size. For caution, we allow
the step size to increase geometrically at each integration step with
only a maximum ratio of 3. Both curves maintain that maximal growth
for several integration steps, indicating that our choice of initial
step size was small enough that any error accumulated during the bootstrapping
phase should be well below the specified desired precision for each
curve. The plateau through the bulk of each integration is followed
by dramatic declines as much greater resolution is needed near the
surface of the sphere.\label{fig:TOV dr vs i}}
\end{figure}

In Figure \ref{fig:TOV dr vs i} we plot the step sizes of each of
our highlighted solutions. We limit the step size to increase by at
most a factor of 3 in each step, in order not to miss any sudden changes
in behavior that might appear. Both curves follow this upper bound
for several steps, indicating that our choice of initial step size
was small enough not to accumulate any significant error during bootstrapping.
The step size then plateaus for the bulk of the radius and then shrinks
near the surface, where much greater resolution is needed.

\begin{figure}[H]
\includegraphics{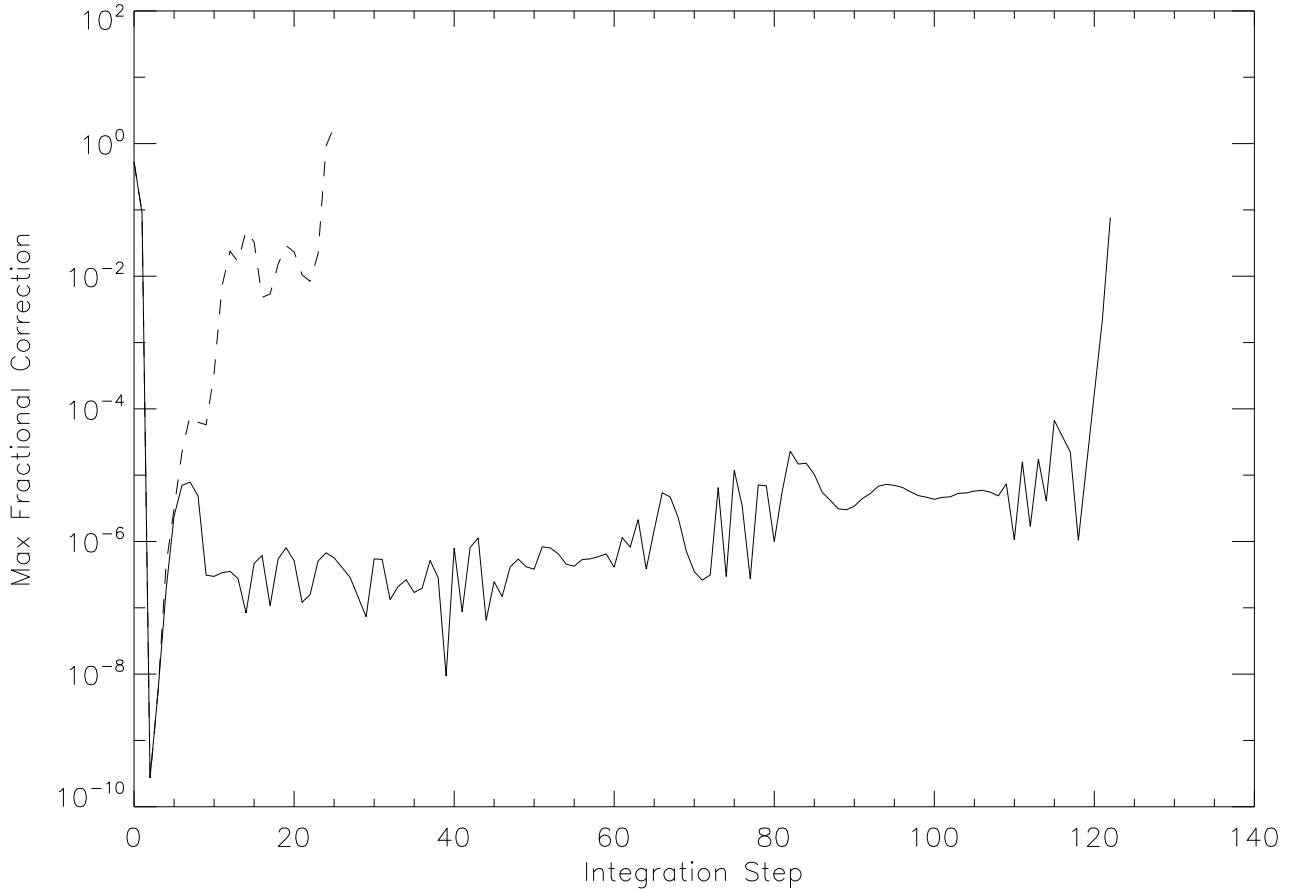}\caption{\emph{AM fractional correction term for high (solid) and low (dashed)
precision ABM solutions to the Tolman-Oppenheimer-Volkoff test problem}
Here we see that both curves maintain stable precision at the desired
level for integration.\label{fig:TOV tol vs i}}
\end{figure}

In Figure \ref{fig:TOV tol vs i} we demonstrate that, after bootstrapping,
the AM correction term remains quite stable throughout most of the
integration steps, increasing only in the final few.

\begin{figure}[H]
\includegraphics{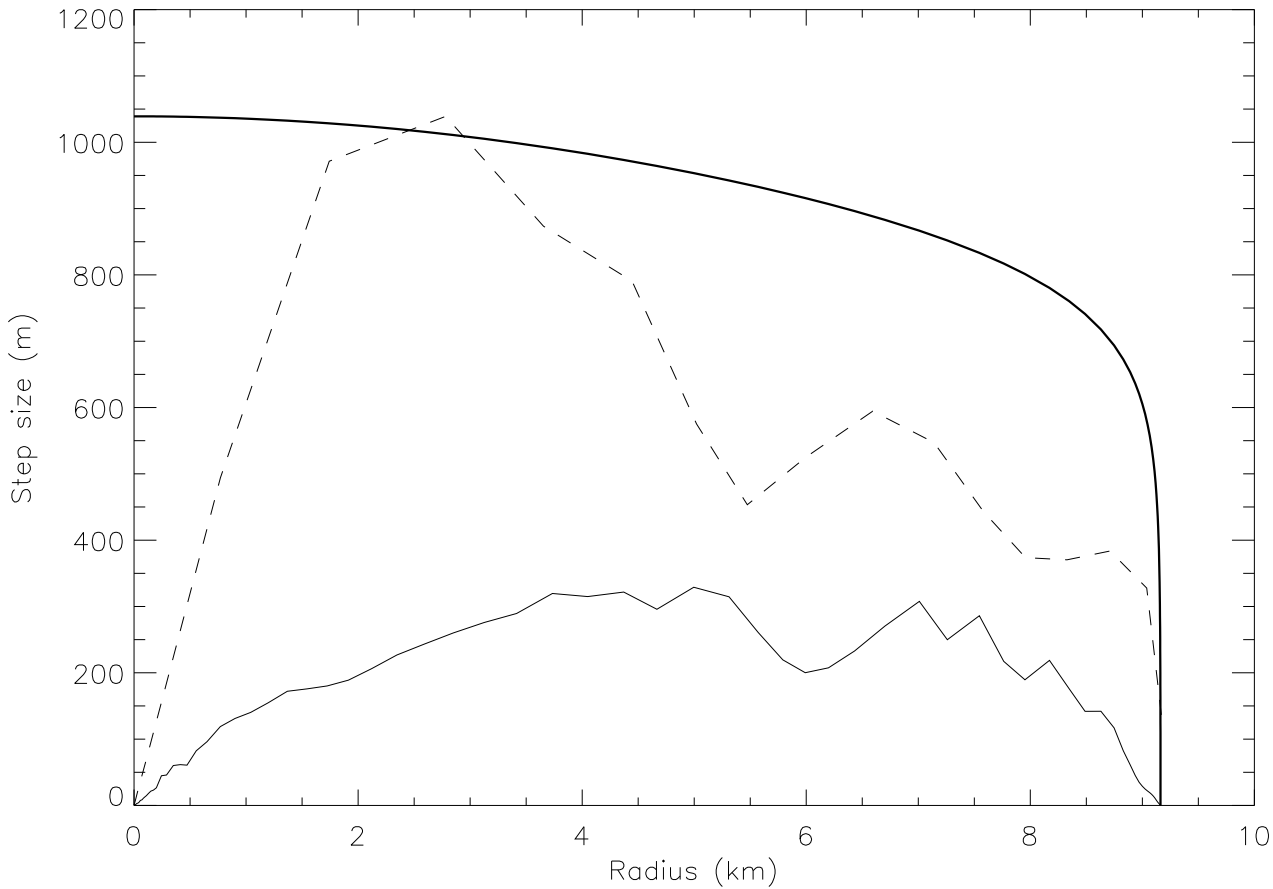}\caption{\emph{Step size vs. r for high (solid) and low (dashed) precision
ABM solutions to the Tolman-Oppenheimer-Volkoff test problem} In order
to clarify why the step size decreases the way it does, we overplot
the log of pressure (in arbitrary units) in a solid, bold line. It
is the dramatic behavior of the pressure near the surface that accounts
for the largest error correction terms, and therefore the decreasing
step size.\label{fig:TOV Step-size-vs.r}}
\end{figure}

In Figure \ref{fig:TOV Step-size-vs.r} we discover the reason for
the increasing size of correction term as shown in Figure \ref{fig:TOV tol vs i}.
We overplot the log of pressure in arbitrary units (bold black curve)
as a function of radius. This shows that the dramatic dropoff in pressure
near the surface is the controlling factor in the need for greater
resolution.

\section{CONCLUSIONS}

We have demonstrated that our integration methods exhibit the expected
order of convergence with various levels of sophistication activated:
\begin{enumerate}
\item Adams-Bashforth-only, fixed-grid, explicit methods
\item Adams-Bashforth-Moulton, fixed-grid, explicit-implicit, predictor-corrector
methods
\item Adams-Bashforth-Moulton, adaptive-grid, explicit-implicit, predictor-corrector
methods.
\end{enumerate}
Furthermore, we have shown that our methods can arrive at high-quality,
robust solutions to real-world research questions in a very small
number of integration steps that require fewer derivative evaluations
than more favored method families.

\section{ACKNOWLEDGEMENTS}

This research was made possible in part by a grant from the Maine
Space Grant Consortium, two Frank H. Todd scholarships, and a Summer
Graduate Research Fellowship and University Graduate Research Assistantship
from the University of Maine. The author would like to thank Chris
Fryer and Kent Budge of Los Alamos National Laboratory for encouragement
to pursue this line of research, Neil F. Comins of the University
of Maine for helpful discussions and editing of this paper, and my
wife Kate and daughter Evangeline for my entire universe.


\begin{thebibliography}{Tolman(1934)}
\bibitem[Bashforth and Adams(1883)]{BA} Bashforth, F. \& Adams, J.
C. 1883, An Attempt to test the Theories of Capillary Action by comparing
the theoretical and measured forms of drops of fluid. With an explanation
of the method of integration employed in constructing the tables which
give the theoretical forms of such drops, (Cambridge University Press)

\bibitem[Courant, Friedrichs, and Lewy(1928)]{Courant}Courant, R.,
Friedrichs, K., \& Lewy, H. 1928, Mathematische Annalen, 100, 1, pp.
32\textendash{}74

\bibitem[Kippenhahn and Weigert(1994)]{Kippenhahn Weigert} Kippenhahn,
R., \& Weigert, A. 1994, Stellar Structure and Evolution (Springer)

\bibitem[Misner, Thorne, and Wheeler(1973)]{MTW} Misner, C., Thorne,
K., \& Wheeler, J. 1973, Gravitation (1973, W.H. Freeman and Company)

\bibitem[Moulton(1926)]{Moulton}Moulton, F. R. 1926, New Methods
in Exterior Ballistics (University of Chicago Press)

\bibitem[Oppenheimer and Volkoff(1939)]{OV} Oppenheimer, J. R. \&
Volkoff, G. M. 1939, Phys. Rev., 55-4, pp. 374\textendash{}381 

\bibitem[Runge(1895)]{Runge}Runge, C. 1895, Math. Ann. 46, pp. 167\textendash{}178

\bibitem[Runge(1901)]{Runge Phenom}Runge, C. 1901, Zeitschrift für
Mathematik und Physik 46: 224\textendash{}243

\bibitem[Schwarzschild(1916)]{Schwarzschild}Schwarzschild, K. 1916,
Proc. Royal Prussian Acad. Sci. meeting on 3 February 1916, p. 189-196

\bibitem[Tolman(1934)]{Tolman}Tolman, R. C. 1934, Proc. Nat. Acad.
Sci. 20 (3): 169\textendash{}176
\end{thebibliography}
\end{document}